\documentclass[12pt,preprint]{aastex}
\newcommand{\sst}{{\it Spitzer Space Telescope}}
\newcommand{\irac}{InfraRed Array Camera}

\newcommand{\eg}{e.g.}

\shorttitle{Mid-Infrared Galaxy Counts}
\shortauthors{Fazio et al.}

\begin{document}

\title{Number Counts At $3 < \lambda < 10 \micron$ from the
                  Spitzer Space Telescope}

\author{
G. G. Fazio\altaffilmark{1}, 
M. L. N. Ashby\altaffilmark{1},
P. Barmby\altaffilmark{1},
J. L. Hora\altaffilmark{1},
J.-S. Huang\altaffilmark{1},
M. A. Pahre\altaffilmark{1},
Z. Wang\altaffilmark{1}, 
S. P. Willner\altaffilmark{1},
R. G. Arendt\altaffilmark{2},
S. H. Moseley\altaffilmark{2},
M. Brodwin\altaffilmark{3}, 
P. Eisenhardt\altaffilmark{3},
Daniel Stern\altaffilmark{3},
E. V. Tollestrup\altaffilmark{4}, \and
E. L. Wright\altaffilmark{5}
}


\altaffiltext{1}{Harvard-Smithsonian Center for Astrophysics, 60
Garden Street, Cambridge, MA 02138; gfazio, leallen, mashby, pbarmby,
jhora, jhuang, mpahre, zwang, swillner@cfa.harvard.edu}

\altaffiltext{2}{Code 685, Goddard Space Flight Center Greenbelt, MD
20771; arendt, moseley@stars.gsfc.nasa.gov}

\altaffiltext{3}{Jet Propulsion Laboratory, 4800 Oak Grove
  Dr. Pasadena, CA 91109; Mark.Brodwin@jpl.nasa.gov,
  peisenhardt@sirtfweb.jpl.nasa.gov, stern@ipac.caltech.edu}

\altaffiltext{4}{Institute for Astronomy, University of Hawaii at
  Hilo, 640 N. A'ohoku Place, \# 209 Hilo, HI 96720;
  tolles@ifa.hawaii.edu}

\altaffiltext{5}{UCLA Dept. of Physics \& Astronomy PO Box 951562 Los
  Angeles, CA 90095-1562; wright@notw33.astro.ucla.edu}

\begin{abstract}
Infrared source counts at wavelengths $3 < \lambda < 10$~\micron\ 
cover more than 10~magnitudes in source brightness, four
orders of magnitude in surface density, and reach an integrated
surface density of $10^5$~sources~deg$^{-2}$.  
At $m<14$~mag, most of the sources are Galactic stars, in
agreement with models.    After removal of Galactic stars, 
galaxy counts are consistent with what few
measurements exist at nearby wavelengths.  At 3.6 and 4.5~\micron,
the galaxy counts follow the expectations of a Euclidean world model
down to $\sim$16~mag and drop below the Euclidean curve for fainter
magnitudes.  Counts at these wavelengths begin to show decreasing
completeness around magnitude 19.5.  At 5.8 and 8~\micron, the counts
relative to a Euclidean world model show a large excess at bright
magnitudes.  This is probably because local galaxies emit strongly in
the aromatic dust (``PAH'') features. The counts at 3.6~\micron\
resolve $<$50\% of the Cosmic Infrared Background at that wavelength.

\end{abstract}



\section{Introduction}

Extragalactic galaxy counts are one of the four ``classical'' tests
of observational cosmology\footnote{
The other three are the redshift-angular size test, the redshift-magnitude
test, and the redshift-surface brightness test. \citet{Peebles}
reviewed the  status of these and more modern tests.}
because they sample the variation in the
volume element with increasing luminosity distance to the source.  Optical
galaxy counts flourished in the last several decades
\citep[e.g.,][]{tyson,hdf} with the advent of charge-coupled devices.
Near-infrared counts arrived in the last decade
\citep[e.g.,][]{d95,yan} with the availability of large format
detectors.  However, the original cosmological goals have been
supplanted by the realization that galaxy evolution plays a major
role in the results.  The optical counts in particular demonstrated
the faint blue galaxy problem---that there were too many faint blue
galaxies compared to predictions based on the local luminosity
function.

In order to understand the effects of galaxy evolution on counts, it
is essential to have counts at a wide range of wavelengths because
there are many types of 
galaxies that emit at different wavelengths over their lifetimes.
For example, normal galaxies alone cannot explain the cosmic infrared background
\citep{hd01}; instead, it is necessary to invoke a population of
dusty, starbursting galaxies (e.g., M82) which re-radiate a substantial portion
of their bolometric luminosity in the far-infrared.  Such galaxies
also exhibit mid-infrared colors redder than normal
galaxies \citep[\eg,][]{Rigopoulou}.  Active galactic nuclei  
have redder colors 
than normal galaxies in the near infrared \citep[\eg, 1.2 to 3.5
\micron][]{Ward}.  Populations of both starbursting
galaxies and AGN have been predicted to be key components of the
extragalactic number counts at wavelengths $>$3\micron.  Some models
predict vastly different numbers of galaxies at 6.7~\micron\ versus
the $K$-band at 2.2\micron. For example, compare the models of
\citet{fran94}, \citet{xu01}, \citet{rr01}, \citet{ms01}, and
\citet{prr96}, which differ by $\sim$1~dex in integrated number
counts at $F_{6.7} \sim 1$~mJy (although they differ less at fainter
fluxes).  Though all these models show reasonable agreement with
the many observations of number counts at 2.2~\micron, 
the disagreement at slightly longer wavelengths
demonstrates that existing data are insufficient to constrain the
models.

Mid-infrared measurements of galaxy counts are exceedingly difficult
to make from the ground because of high thermal background emission
and the limited wavelength regions that the atmosphere is
transparent.  Nonetheless, \citet{hogg} detected nine extragalactic
sources to a limit of 17.5~mag (33~$\mu$Jy) at 3.2\micron .
Observations at mid-infrared wavelengths are better made from space,
where the thermal background is low and stable.  The {\it Infrared
Space Observatory} mission led to the first observations
of galaxy counts at 6.75\micron\ \citep{Oliver1997}, reaching
15.8~mag (40~$\mu$Jy) in the Hubble Deep Field.  Similar observations (or
up to 0.2~mag deeper) were reported by \citet{Taniguchi1997} for the
Lockman Hole, \citet{Flores1999} for the CFRS 1415+52 field (part of
the EGS, discussed below), and \citet{Oliver2002} for the Hubble Deep
Field South.  \citet{Altieri1999} observed a lensing cluster (A2390)
to about the same depth on the sky but detected intrinsically fainter
sources because of the lens amplification.  \citet{sato03} reached
sources as faint as 17.8~mag (6~$\mu$Jy) in the SSA~13 region.
\citet{elais} took a different approach, surveying to only 12.3~mag
(1~mJy) but covering a large area of 6.5~deg$^2$.

Here, we report the first source counts at $3 < \lambda <
10$~\micron\ made with the \irac\ (IRAC) instrument on the \sst. With
the low background available in space, the instrument can reach faint
limiting fluxes in modest exposure times.  In fact, IRAC can reach
similar depths (19th magnitude) as the 10~m W.\ M.\ Keck Telescope
with the NIRC2 camera at $\sim$~3.6~\micron\ in only one-hundredth of
the exposure time and does so with a far larger field of view:
$5'\times5'$ versus $40''\times40''$.

In order to cover a wide range of magnitudes and counts, we combine
data from three survey fields of complementary area and depth.  We
first present the raw source counts, which give the first
comprehensive picture of the infrared sky at these wavelengths and
sensitivities.  We then subtract the Galactic stellar contribution to
derive the extragalactic source counts.  Quantitative data in this
paper are given in instrumental magnitudes relative to Vega, the
units least dependent on calibration uncertainties.  The flux
density of Vega in the IRAC bandpasses is
$(277.5,179.5,116.6,63.1)$~Jy \citep{irac}.

\section{Observations and data analysis}

The deepest image used here \citep{iocdeep} is of a $5'\times10'$
field surrounding 
the QSO HS 1700+6416 (hereafter QSO1700).\footnote{
This field
was observed in 2003 October as Spitzer program id (pid) $=$ 620.} 
Heavily
dithered coverage allowed removal of spurious sources such as cosmic
rays, instrumental artifacts due to bright stars, and the wings of
bright stars.  The resulting average exposure time at each point in
the field was 7.8 hours.  Object detection used SExtractor
\citep{ba96} with the detection threshold set to 2.5$\sigma$ and the
minimum area 5 pixels (3.6, 4.5~$\mu$m) or 7 pixels (5.8,
8.0~$\mu$m).  Objects without a central peak (bright star wings) or
sets of objects that appeared in a line (`muxbleed' or `banding'
artifacts) were removed. The 5$\sigma$ depth reached was 22.0, 21.5,
20.4, 19.6~mag (0.4, 0.5, 0.8, and 0.9 $\mu$Jy) respectively at the
four IRAC wavelengths 3.6, 4.5, 5.8, and 
8.0~\micron\ \citep{iocdeep}.
Custom IDL software was used to measure magnitudes of detected objects 
in $1\farcs5$ and $3\farcs0$ radius circular apertures with
sky annuli of $25\arcsec < r < 35\arcsec$.  A correction to a
$12\farcs2$ radius aperture (the standard aperture for observations
of IRAC calibration stars) was derived using curves of growth based
on the in-flight IRAC point spread function.\footnote{
The correction factors used were
$-$0.52, $-$0.55, $-$0.74, and $-$0.85~mag for $r=1\farcs5$  and
$-$0.15, $-$0.14, $-$0.20, $-$0.35~mag for $r=3\farcs0$ for the four
IRAC wavelengths.}

The intermediate depth image covers $0\fdg17\times2\degr$ in the
Extended Groth Strip (hereafter EGS).  Data processing and photometry
were the same as for the QSO1700 field.\footnote{The EGS field was
observed in 2003 December as part of the Spitzer Guaranteed Time
Observer (GTO) project ``The IRAC Deep Survey,'' pid $=$ 8.  Coverage
of this field is expected to be repeated later in the Spitzer
mission. The DEEP2 survey \citep{deep2} has produced complementary
visible and near-infrared data for this field.}  Depth of coverage
was 26 frames of 200-s each reaching 5$\sigma$ limits of 21.5, 21.0,
18.5, 17.9~mag (0.7, 0.7, 4.6, 4.4~$\mu$Jy) at the four IRAC
wavelengths.

The widest area survey covers $\sim 3\degr\times3\degr$ in the Bo\"otes
region of the ``NOAO Deep-Wide Survey'' \citep{ndwfs}.
Depth of
coverage is three 30-s IRAC frames at each point.\footnote{
IRAC data were taken in 2004 January as part of the GTO project ``IRAC
shallow survey,'' pid $=$ 30.}  Because of the
minimal dithering, all sky positions having fewer than three
individual images in agreement were removed from the mosaic to avoid
having chance coincidences of cosmic rays ``detected'' as objects.
The limiting magnitudes for 5$\sigma$ detection are 18.4, 17.7, 15.5,
and 14.8 (12, 15, 74, 76~$\mu$Jy) at the four IRAC wavelengths \citep{shallow}.
Source magnitudes were measured with SExtractor.  The bright end
number counts, which mostly refer to stars, are limited by saturation
at magnitudes of 10.0, 9.8, 7.5, 7.7 (28, 22, 117, 52~mJy).

\begin{figure}
\epsscale{1.00}
\plotone{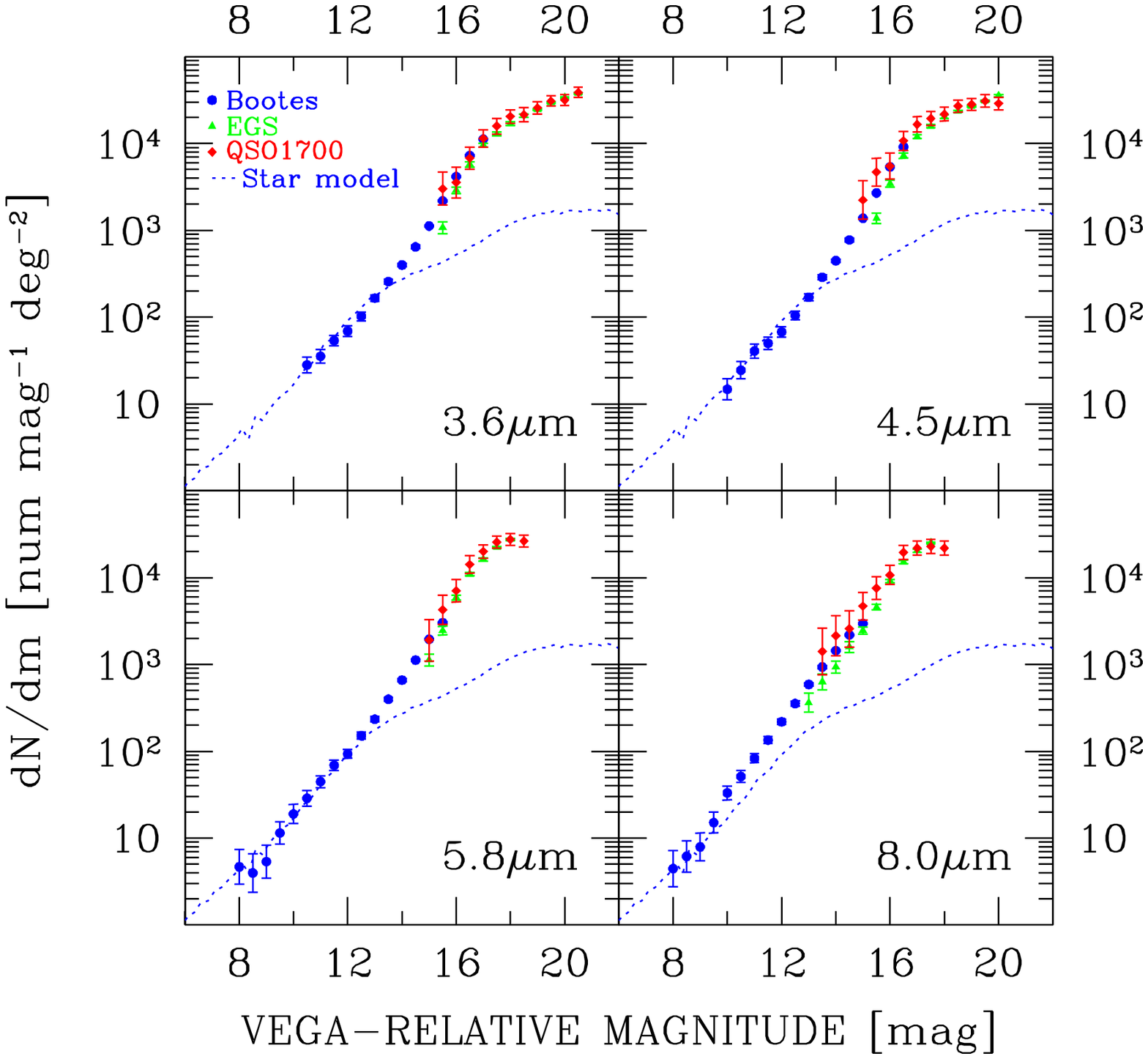}
\caption{ Differential number counts in the four IRAC bandpasses
including all sources, stars and galaxies.  Counts in the three
fields are indicated by different colors.  Only bins containing at
least 10 sources are included.  The dashed line shows stellar number
counts predicted by the Faint Source Model as used by the DIRBE
mission data analysis \citep{dirbe} for the Bo\"otes field.  
Stars dominate at the bright
end of the counts but are a minor contributor at the faint end.
\label{diffcounts} }
\end{figure}

Table~\ref{numcts} (columns 2, 6, and 11) and Figure~\ref{diffcounts}
show the resulting number counts for each field.


In order to use the number counts to study galaxy evolution, stars
must be subtracted.  At bright magnitudes (Bo\"otes field), the
spatial extent of brighter galaxies is big enough to separate them
from stars using the difference between aperture magnitudes and
``magauto'' (which is roughly an isophotal magnitude) generated by
SExtractor.  Near the bright limit of the survey, the distant wings
of the stellar PSF are bright enough to be seen, and bright stars are
classed as galaxies by the automated software.  Therefore the few
sources in this brightness range were examined and classified
visually.  At fainter magnitudes, $m \ga 14.5$~mag, morphology can no
longer easily separate stars and galaxies at our image resolution of
$\geq$1\farcs7 FWHM\null.  The stellar contribution to the total
source counts at these magnitudes, however, is small, so their number
can be estimated statistically.  Table~\ref{numcts} shows
measured star counts (column 4) for the Bo\"otes field and 
model star counts (columns 8 and 13) for the two deeper fields.
The model \citep{dirbe} is similar to one by \citet{Wainscoat1992}
but with additional geometric details and extended source colors.  The model
matches the unresolved Galactic emission seen by DIRBE but may not
accurately predict actual star counts in all lines of sight.  Despite
that, for $m>16.5$~mag, stars are a negligible fraction of the total
counts.  In the troublesome region $14.5<m<16.5$~mag, results for
different fields and methods agree, and the galaxy counts do not
change slope, but the extragalactic counts may be less accurate than
in other ranges.

At the faintest levels, all surveys fail to detect some objects brighter
than the ostensible limiting magnitude.  This incompleteness is
estimated by inserting artificial sources into the sky images and
finding their recovery rate as a function of magnitude.  For the EGS
and QSO1700 fields, we used the usual `Monte Carlo' method: scaling
images of objects in the field to fainter fluxes, inserting them in
the images in sets of 200, re-running SExtractor, and counting the
number of artificial objects recovered. Objects were considered to be
recovered if they were detected within 1.5~pixels and 0.5~mag of
their input positions and magnitudes.  Source counts were truncated
at the faint end when the incompleteness exceeded 50\%.  The results
are shown in Table~\ref{numcts} (columns 9 and 14).

The incompleteness in the 3.6 and 4.5~\micron\ images does not show
the usual rapid increase near the magnitude limit of the images but rather
shows a gradual increase that begins at bright magnitudes
$([3.6],[4.5]) \approx (17.5,17.0)$.\footnote{
Here and in other IRAC papers, the notation [$\lambda$] is used to
mean magnitude at wavelength $\lambda$~\micron.}
Furthermore, the
incompleteness curves at these two wavelengths are very similar for
both the EGS and QSO1700 fields, despite the differing exposure times by
a factor of $>$5.  Both of these features of the data are
consistent with the images being affected by source confusion at
their faint limits.  For this reason, we choose to truncate the
source counts at a relatively bright magnitude $[3.6] = [4.5] =
20.5$~mag.  This corresponds to integral counts of $\sim$36~beams/source. 
We caution against any over-interpretation of our results
at the faintest magnitudes due to the complications arising from the
source confusion.  An improved analysis---using more sophisticated
image combination algorithms, source detection, and more extensive
Monte Carlo incompleteness simulations---will be deferred to a future
contribution.

The incompleteness factors at 5.8 and 8~\micron\ show sharper
declines near the faint limits at these wavelengths and significant
improvement in the QSO1700 field compared to the EGS field. Evidently
the images at these wavelengths are still not deep enough for
confusion to be a strong factor.  This is consistent with the
integral source density of $\sim$60 beams/source to magnitude 18.

Incompleteness corrections were not calculated for the Bo\"otes field
because its faint flux limit overlaps the other surveys.  The limit
indicated in Table~\ref{numcts} is where the Bo\"otes field
counts drop below 95\% of the counts expected from the EGS field.


Figure~\ref{numcts} shows that Galactic stars dominate the
counts at the bright end but become much less important at $m>
14.5$~mag.  Approximately $(4.8 \times 10^4,7.3 \times 10^3,2.8
\times 10^2)$ sources were catalogued at $[3.6] <
(17.25,20.75,20.75)$~mag in the Bo\"otes, EGS, and QSO1700 fields,
respectively.  The incompleteness-corrected integrated source counts
reach surface densities of $\sim (1.4 \times 10^5,1.2 \times 10^5,7.1
\times 10^4,6.8 \times 10^4)$~deg$^{-2}$ at the wavelengths of
$(3.6,4.5,5.8,8.0)$ respectively.  The QSO1700 field contains two
galaxy clusters at 
redshifts of 0.25 and 0.44.  These may contribute to increased counts
in the $15.5 < m < 17$ range for this field.

\section{Galaxy Counts}

\begin{figure}
\epsscale{0.90}
\plotone{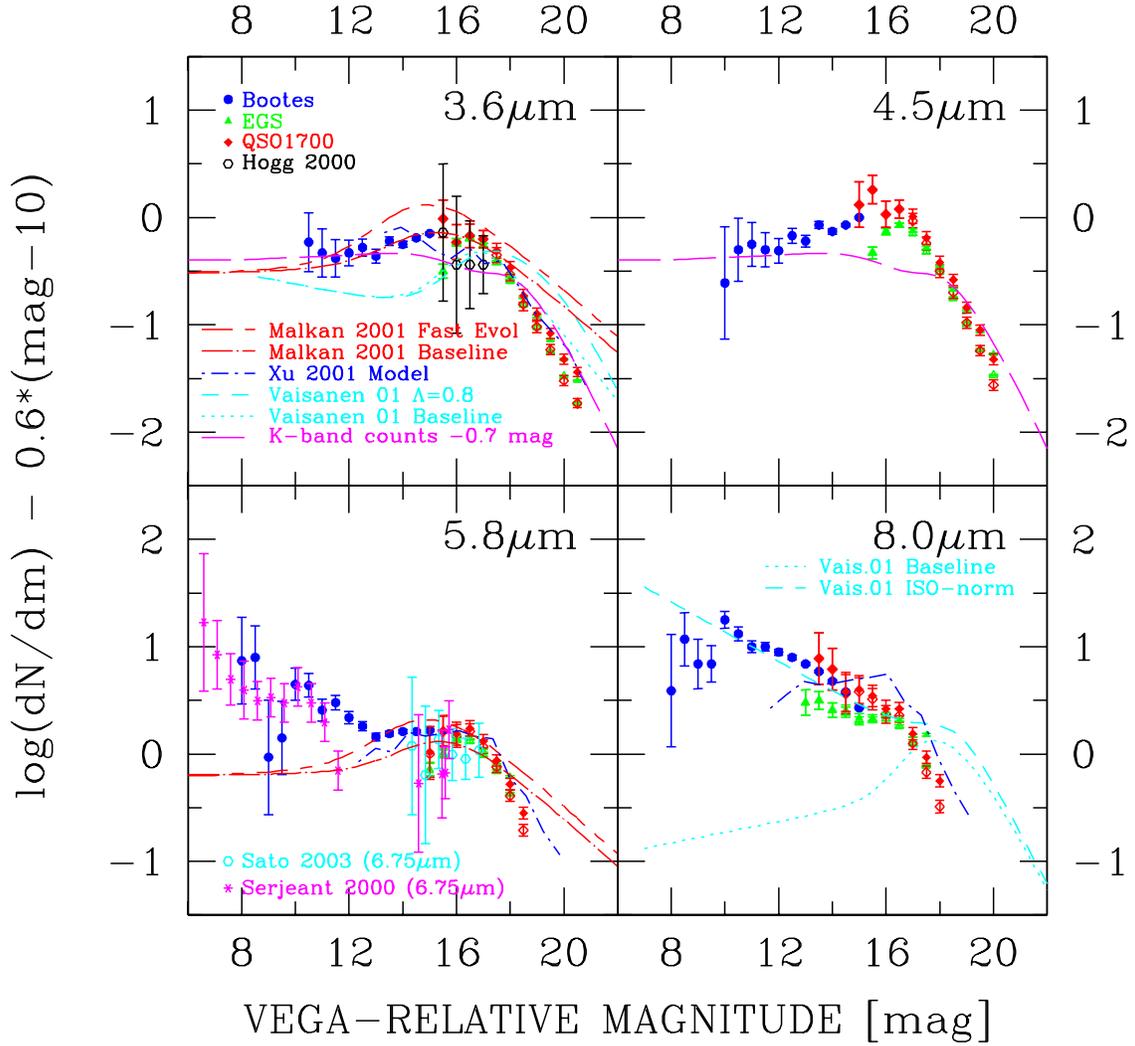}
\caption{ Extragalactic differential number counts in the four IRAC
bandpasses normalized by the expected growth rate in a Euclidean
Universe, i.e., $N(m)\times 10^{[-0.6(m-10)]}$.  Open symbols
are raw galaxy counts; filled symbols have the derived incompleteness
correction applied.  Stars were removed from the Bo\"otes data using
a morphological criterion and from the EGS and QSO1700 data
statistically using the Galactic Faint Source Model
(\citealt{dirbe}; see text).  The 3.6~\micron\ panel shows comparison
data from \citet[][$3.2\mu$m]{hogg} and the $K$-band number counts
model from \citet{kochanek01}, which is a good representation of the
$K$-band data in this brightness range.  The \citeauthor{kochanek01}
curve has been shifted by an arbitrary -0.7~mag to better match the
3.6~\micron\ counts; this value could be interpreted as the average
$K-[3.6]$ color for the galaxies contributing to the counts.  The
5.8~\micron\ panel contains comparison 6.75~\micron\ counts from
\citet{elais} and \citet{sato03}.  Also plotted are  models from
\citet[3.6 and 5.8~\micron]{ms01}, 
\citet[3.6, 5.8, and 8~\micron]{xu01}, and 
\citet[3.6 and 8~\micron]{vtf01}.
\label{diffeuclidean} }
\end{figure}

Figure~\ref{diffeuclidean} shows our best estimate of the
extragalactic number counts with stars subtracted and completeness
corrections included.  Some of the previous galaxy counts at nearby
wavelengths are also shown as are some simple models.  There is
general agreement with previous observations, although wavelength
differences may affect the results.  For example, the ISO LW2 filter
overlaps considerably with both the 5.8 and 8.0~\micron\ bandpasses
of IRAC, but it shows consistency only with our 5.8~\micron\ counts.
This is most likely a result of the differing long wavelength cutoffs
of the filters (8.5~\micron\ for LW2 versus 9.5~\micron\ for the IRAC
$8 \mu$m filter).


The 3.6 and 4.5~\micron\ observations are best compared with
ground-based $K$-band (2.2~\micron) counts. There are many surveys in
the literature, but the curve from \citet{kochanek01} represents all
the data. The IRAC counts in the two bands are consistent with the
prior observations 
with only a mean offset of $(K-[3.6]) \sim 0.7$~mag being required to
bring them into rough agreement at $[3.6] < 17$~mag.  
Models that predict the 3.6~\micron\ counts to differ
markedly from the $K$-band counts can therefore be rejected.
The various models
plotted in Figure~\ref{diffeuclidean} demonstrate that there is no
single model to date that can match the galaxy counts throughout the
observed flux ranges and in all bandpasses simultaneously.

\section{Discussion}

The galaxy counts at 3.6 and 4.5~\micron\ appear flat at $m \leq
16$~mag, consistent with a Euclidean world model.  The $5.8$ and
8.0~\micron\ counts, however, show a steep drop from $8<m<16$~mag. We
attribute the drop to substantial contribution to the fluxes from the
strong aromatic feature emission at 6.2 and 7.7~\micron\
\citep[e.g.,][]{lu03}.  The brightest galaxies tend to be local
ones, and the emission features are within the IRAC bands.  Fainter galaxies,
on the other hand, tend to be at higher redshift, and at $z \geq
0.23$ the 7.7~\micron\ feature is leaving the 8.0~\micron\ IRAC band.
In effect, these bands have a strong positive ``K-correction''
(decreasing intrinsic flux density) as
redshift increases, as suggested by \citet{Aussel1999} for the ISO
6.75~\micron\ filter.  The slopes have different shapes in the two
bands, presumably because of the larger bandwidth of the 8~\micron\
filter and the greater strength of the 7.7~\micron\ feature relative
to the 6.2~\micron\ feature.

With these number counts, we can estimate the contribution of
IRAC-detected galaxies to the cosmic infrared background (CIRB).  The
integrated galaxy counts (weighted according to uncertainties) in
Table~\ref{numcts} correspond to 5.4, 3.5, 3.6, and
2.6~nW~m$^{-2}$~sr$^{-1}$ at the four IRAC wavelengths.  The
3.6~\micron\ surface brightness is  $<$50\% of the estimated
12~nW~m$^{-2}$~sr$^{-1}$ CIRB at this wavelength
\citep{Wright2000}. At longer wavelengths (4.5 -- 8~\micron), no
reliable direct measurements of the EBL exist because of the
difficulty in accurately accounting for the zodiacal and Galactic
foreground emission, which brighten rapidly at $\lambda
\gtrsim 3.5$~\micron.  It may prove difficult to construct a model
consistent with these resolved galaxy counts while producing
enough far infrared flux to match the cosmic far-infrared background.


\section{Summary}

Source counts at $3 < \lambda < 10 \mu$m taken with the IRAC
instrument on the \sst\ demonstrate:

\begin{enumerate}

\item Galaxy counts follow the Euclidean expectation at brighter
fluxes at $3.6$ and $4.5 \mu$m;

\item Confusion begins to set in around 19.5~mag for the 3.6 and
4.5~\micron\ images and present source extraction technique, but
there is no evidence of confusion down to 18.5~mag at 5.8~\micron\
and 18.0~mag at 8~\micron;

\item Bright galaxy counts at 5.8 and 8.0~\micron\ are more numerous
than the Euclidean expectation, most likely due to PAH emission lines
from low-redshift, star-forming galaxies;

\item No existing model matches the galaxy counts throughout the
observed flux range and simultaneously in all the IRAC bandpasses.

\end{enumerate}

\acknowledgments

This work is based on observations made with the \sst, which is
operated by the Jet Propulsion Laboratory, California Institute of
Technology under NASA contract 1407. Support for this work was
provided by NASA through Contract Number  1256790 issued by
JPL/Caltech.
M.A.P. acknowledges NASA/LTSA grant \# NAG5-10777.

Facilities:  \facility{Spitzer(IRAC)}.



\clearpage

\clearpage

\clearpage






%


\begin{deluxetable}{lcccccccccccccc}
\tabletypesize{\scriptsize}
\rotate
\tablecaption{IRAC Differential Number Counts  \label{numcts}}
\tablewidth{0pt}
\tablehead{
& 
\colhead{Bo\"otes} & & & &
\colhead{EGS} & & & & &
\colhead{QSO1700} &  \\
\colhead{Magnitude} & 
\colhead{Total} & 
\colhead{$\pm$} &
\colhead{Stars\tablenotemark{a}} & 
\colhead{Galaxies} & 
\colhead{Total} & 
\colhead{$\pm$} & 
\colhead{Stars\tablenotemark{b}} & 
\colhead{Comp.} &
\colhead{Galaxies\tablenotemark{c}} & 
\colhead{Total} & 
\colhead{$\pm$} & 
\colhead{Stars\tablenotemark{b}} & 
\colhead{Comp.} &
\colhead{Galaxies\tablenotemark{c}} 
 \\
\colhead{1} &
\colhead{2} &
\colhead{3} &
\colhead{4} &
\colhead{5} &
\colhead{6} &
\colhead{7} &
\colhead{8} &
\colhead{9} &
\colhead{10} &
\colhead{11} &
\colhead{12} &
\colhead{13} &
\colhead{14} &
\colhead{15} 
}
\startdata

\multicolumn{9}{l}{$\lambda = 3.6 \mu$m} \\ 
10.5 &    1.45 &    0.09 &    1.43 &    0.07 & \nodata & \nodata & \nodata & \nodata & \nodata & \nodata & \nodata & \nodata & \nodata &   \nodata  \\
11.0 &    1.55 &    0.08 &    1.52 &    0.27 & \nodata & \nodata & \nodata & \nodata & \nodata & \nodata & \nodata & \nodata & \nodata &   \nodata  \\
11.5 &    1.73 &    0.06 &    1.71 &    0.52 & \nodata & \nodata & \nodata & \nodata & \nodata & \nodata & \nodata & \nodata & \nodata &   \nodata  \\
12.0 &    1.84 &    0.06 &    1.80 &    0.87 & \nodata & \nodata & \nodata & \nodata & \nodata & \nodata & \nodata & \nodata & \nodata &   \nodata  \\
12.5 &    2.01 &    0.05 &    1.93 &    1.22 & \nodata & \nodata & \nodata & \nodata & \nodata & \nodata & \nodata & \nodata & \nodata &   \nodata  \\
13.0 &    2.22 &    0.04 &    2.14 &    1.44 & \nodata & \nodata & \nodata & \nodata & \nodata & \nodata & \nodata & \nodata & \nodata &   \nodata  \\
13.5 &    2.41 &    0.03 &    2.27 &    1.88 & \nodata & \nodata & \nodata & \nodata & \nodata & \nodata & \nodata & \nodata & \nodata &   \nodata  \\
14.0 &    2.60 &    0.02 &    2.41 &    2.15 & \nodata & \nodata & \nodata & \nodata & \nodata & \nodata & \nodata & \nodata & \nodata &   \nodata  \\
14.5 &    2.81 &    0.02 &    2.51 &    2.51 & \nodata & \nodata & \nodata & \nodata & \nodata & \nodata & \nodata & \nodata & \nodata &   \nodata  \\
15.0 &    3.05 &    0.01 &    2.61 &    2.85 & \nodata & \nodata & \nodata & \nodata & \nodata & \nodata & \nodata & \nodata & \nodata &   \nodata  \\
15.5 &    3.34 &    0.01 & \nodata & \nodata &    3.03 &    0.07 &    2.66 &    1.00 &    2.80 &    3.48 &    0.19 &    3.03 &    1.00 &      3.29  \\
16.0 &    3.62 &    0.01 & \nodata & \nodata &    3.46 &    0.04 &    2.71 &    1.00 &    3.37 &    3.55 &    0.18 &    3.09 &    1.00 &      3.37  \\
16.5 &    3.86 &    0.01 & \nodata & \nodata &    3.76 &    0.03 &    2.78 &    1.00 &    3.71 &    3.83 &    0.13 &    3.14 &    1.00 &      3.73  \\
17.0 &    4.05 &    0.00 & \nodata & \nodata &    3.98 &    0.02 &    2.85 &    1.00 &    3.94 &    4.06 &    0.10 &    3.20 &    1.00 &      4.00  \\
17.5 & \nodata & \nodata & \nodata & \nodata &    4.11 &    0.02 &    2.91 &    0.94 &    4.11 &    4.20 &    0.09 &    3.23 &    0.91 &      4.19  \\
18.0 & \nodata & \nodata & \nodata & \nodata &    4.23 &    0.02 &    2.96 &    0.89 &    4.26 &    4.31 &    0.08 &    3.26 &    0.86 &      4.33  \\
18.5 & \nodata & \nodata & \nodata & \nodata &    4.32 &    0.02 &    3.00 &    0.86 &    4.36 &    4.33 &    0.08 &    3.28 &    0.83 &      4.37  \\
19.0 & \nodata & \nodata & \nodata & \nodata &    4.39 &    0.01 &    3.02 &    0.79 &    4.48 &    4.41 &    0.07 &    3.27 &    0.76 &      4.50  \\
19.5 & \nodata & \nodata & \nodata & \nodata &    4.47 &    0.01 &    3.01 &    0.77 &    4.56 &    4.49 &    0.06 &    3.24 &    0.69 &      4.62  \\
20.0 & \nodata & \nodata & \nodata & \nodata &    4.53 &    0.01 &    3.01 &    0.69 &    4.68 &    4.50 &    0.06 &    3.20 &    0.63 &      4.68  \\
20.5 & \nodata & \nodata & \nodata & \nodata &    4.58 &    0.01 &    3.02 &    0.61 &    4.78 &    4.59 &    0.06 &    3.17 &    0.52 &      4.86  \\

\multicolumn{9}{l}{$\lambda = 4.5 \mu$m} \\ 
10.0 &    1.17 &    0.12 &    1.16 &   -0.61 & \nodata & \nodata & \nodata & \nodata & \nodata & \nodata & \nodata & \nodata & \nodata &   \nodata  \\
10.5 &    1.39 &    0.10 &    1.38 &   -0.00 & \nodata & \nodata & \nodata & \nodata & \nodata & \nodata & \nodata & \nodata & \nodata &   \nodata  \\
11.0 &    1.61 &    0.08 &    1.58 &    0.35 & \nodata & \nodata & \nodata & \nodata & \nodata & \nodata & \nodata & \nodata & \nodata &   \nodata  \\
11.5 &    1.70 &    0.07 &    1.66 &    0.60 & \nodata & \nodata & \nodata & \nodata & \nodata & \nodata & \nodata & \nodata & \nodata &   \nodata  \\
12.0 &    1.83 &    0.06 &    1.78 &    0.89 & \nodata & \nodata & \nodata & \nodata & \nodata & \nodata & \nodata & \nodata & \nodata &   \nodata  \\
12.5 &    2.02 &    0.05 &    1.92 &    1.33 & \nodata & \nodata & \nodata & \nodata & \nodata & \nodata & \nodata & \nodata & \nodata &   \nodata  \\
13.0 &    2.23 &    0.04 &    2.12 &    1.58 & \nodata & \nodata & \nodata & \nodata & \nodata & \nodata & \nodata & \nodata & \nodata &   \nodata  \\
13.5 &    2.46 &    0.03 &    2.27 &    2.03 & \nodata & \nodata & \nodata & \nodata & \nodata & \nodata & \nodata & \nodata & \nodata &   \nodata  \\
14.0 &    2.65 &    0.02 &    2.42 &    2.27 & \nodata & \nodata & \nodata & \nodata & \nodata & \nodata & \nodata & \nodata & \nodata &   \nodata  \\
14.5 &    2.89 &    0.02 &    2.53 &    2.63 & \nodata & \nodata & \nodata & \nodata & \nodata & \nodata & \nodata & \nodata & \nodata &   \nodata  \\
15.0 &    3.14 &    0.01 &    2.58 &    3.00 & \nodata & \nodata & \nodata & \nodata & \nodata &    3.35 &    0.22 &    2.96 &    1.00 &      3.12  \\
15.5 &    3.43 &    0.01 & \nodata & \nodata &    3.14 &    0.06 &    2.66 &    1.00 &    2.97 &    3.67 &    0.16 &    3.03 &    1.00 &      3.56  \\
16.0 &    3.73 &    0.01 & \nodata & \nodata &    3.54 &    0.04 &    2.71 &    1.00 &    3.47 &    3.74 &    0.15 &    3.09 &    1.00 &      3.63  \\
16.5 &    3.96 &    0.01 & \nodata & \nodata &    3.86 &    0.03 &    2.78 &    1.00 &    3.83 &    4.03 &    0.11 &    3.14 &    1.00 &      3.98  \\
17.0 & \nodata & \nodata & \nodata & \nodata &    4.08 &    0.02 &    2.85 &    0.93 &    4.08 &    4.22 &    0.09 &    3.20 &    0.92 &      4.21  \\
17.5 & \nodata & \nodata & \nodata & \nodata &    4.20 &    0.02 &    2.91 &    0.91 &    4.22 &    4.29 &    0.08 &    3.23 &    0.89 &      4.31  \\
18.0 & \nodata & \nodata & \nodata & \nodata &    4.31 &    0.02 &    2.96 &    0.89 &    4.35 &    4.34 &    0.08 &    3.26 &    0.84 &      4.38  \\
18.5 & \nodata & \nodata & \nodata & \nodata &    4.37 &    0.01 &    3.00 &    0.83 &    4.43 &    4.43 &    0.07 &    3.28 &    0.77 &      4.52  \\
19.0 & \nodata & \nodata & \nodata & \nodata &    4.43 &    0.01 &    3.02 &    0.77 &    4.53 &    4.45 &    0.07 &    3.27 &    0.72 &      4.56  \\
19.5 & \nodata & \nodata & \nodata & \nodata &    4.49 &    0.01 &    3.01 &    0.67 &    4.64 &    4.49 &    0.07 &    3.24 &    0.65 &      4.65  \\
20.0 & \nodata & \nodata & \nodata & \nodata &    4.55 &    0.01 &    3.01 &    0.64 &    4.73 &    4.46 &    0.07 &    3.20 &    0.57 &      4.68  \\

\multicolumn{9}{l}{$\lambda = 5.8 \mu$m} \\ 
 8.0 &    0.67 &    0.20 &    0.63 &   -0.33 & \nodata & \nodata & \nodata & \nodata & \nodata & \nodata & \nodata & \nodata & \nodata &   \nodata  \\
 8.5 &    0.60 &    0.22 &    0.60 &    0.00 & \nodata & \nodata & \nodata & \nodata & \nodata & \nodata & \nodata & \nodata & \nodata &   \nodata  \\
 9.0 &    0.73 &    0.19 &    0.71 &   -0.63 & \nodata & \nodata & \nodata & \nodata & \nodata & \nodata & \nodata & \nodata & \nodata &   \nodata  \\
 9.5 &    1.06 &    0.13 &    1.03 &   -0.15 & \nodata & \nodata & \nodata & \nodata & \nodata & \nodata & \nodata & \nodata & \nodata &   \nodata  \\
10.0 &    1.28 &    0.11 &    1.16 &    0.65 & \nodata & \nodata & \nodata & \nodata & \nodata & \nodata & \nodata & \nodata & \nodata &   \nodata  \\
10.5 &    1.46 &    0.09 &    1.31 &    0.94 & \nodata & \nodata & \nodata & \nodata & \nodata & \nodata & \nodata & \nodata & \nodata &   \nodata  \\
11.0 &    1.65 &    0.07 &    1.54 &    1.01 & \nodata & \nodata & \nodata & \nodata & \nodata & \nodata & \nodata & \nodata & \nodata &   \nodata  \\
11.5 &    1.84 &    0.06 &    1.66 &    1.38 & \nodata & \nodata & \nodata & \nodata & \nodata & \nodata & \nodata & \nodata & \nodata &   \nodata  \\
12.0 &    1.97 &    0.05 &    1.78 &    1.54 & \nodata & \nodata & \nodata & \nodata & \nodata & \nodata & \nodata & \nodata & \nodata &   \nodata  \\
12.5 &    2.18 &    0.04 &    1.97 &    1.76 & \nodata & \nodata & \nodata & \nodata & \nodata & \nodata & \nodata & \nodata & \nodata &   \nodata  \\
13.0 &    2.37 &    0.03 &    2.15 &    1.96 & \nodata & \nodata & \nodata & \nodata & \nodata & \nodata & \nodata & \nodata & \nodata &   \nodata  \\
13.5 &    2.60 &    0.02 &    2.31 &    2.29 & \nodata & \nodata & \nodata & \nodata & \nodata & \nodata & \nodata & \nodata & \nodata &   \nodata  \\
14.0 &    2.82 &    0.02 &    2.41 &    2.61 & \nodata & \nodata & \nodata & \nodata & \nodata & \nodata & \nodata & \nodata & \nodata &   \nodata  \\
14.5 &    3.05 &    0.01 &    2.50 &    2.91 & \nodata & \nodata & \nodata & \nodata & \nodata & \nodata & \nodata & \nodata & \nodata &   \nodata  \\
15.0 &    3.29 &    0.01 &    2.48 &    3.22 &    3.05 &    0.07 &    2.60 &    0.98 &    2.86 &    3.28 &    0.24 &    2.96 &    0.96 &      3.02  \\
15.5 &    3.48 &    0.01 & \nodata & \nodata &    3.39 &    0.05 &    2.66 &    0.96 &    3.31 &    3.63 &    0.17 &    3.03 &    0.96 &      3.52  \\
16.0 & \nodata & \nodata & \nodata & \nodata &    3.77 &    0.03 &    2.71 &    0.95 &    3.75 &    3.85 &    0.13 &    3.09 &    0.94 &      3.79  \\
16.5 & \nodata & \nodata & \nodata & \nodata &    4.04 &    0.02 &    2.78 &    0.92 &    4.05 &    4.15 &    0.10 &    3.14 &    0.92 &      4.14  \\
17.0 & \nodata & \nodata & \nodata & \nodata &    4.21 &    0.02 &    2.85 &    0.92 &    4.23 &    4.30 &    0.08 &    3.20 &    0.89 &      4.32  \\
17.5 & \nodata & \nodata & \nodata & \nodata &    4.35 &    0.01 &    2.91 &    0.82 &    4.42 &    4.41 &    0.07 &    3.23 &    0.87 &      4.44  \\
18.0 & \nodata & \nodata & \nodata & \nodata &    4.44 &    0.01 &    2.96 &    0.69 &    4.59 &    4.44 &    0.07 &    3.26 &    0.78 &      4.52  \\
18.5 & \nodata & \nodata & \nodata & \nodata & \nodata & \nodata & \nodata & \nodata & \nodata &    4.42 &    0.07 &    3.28 &    0.69 &      4.55  \\

\multicolumn{9}{l}{$\lambda = 8.0 \mu$m} \\ 
 8.0 &    0.65 &    0.21 &    0.62 &   -0.61 & \nodata & \nodata & \nodata & \nodata & \nodata & \nodata & \nodata & \nodata & \nodata &   \nodata  \\
 8.5 &    0.79 &    0.18 &    0.67 &    0.17 & \nodata & \nodata & \nodata & \nodata & \nodata & \nodata & \nodata & \nodata & \nodata &   \nodata  \\
 9.0 &    0.90 &    0.16 &    0.79 &    0.24 & \nodata & \nodata & \nodata & \nodata & \nodata & \nodata & \nodata & \nodata & \nodata &   \nodata  \\
 9.5 &    1.18 &    0.12 &    1.07 &    0.54 & \nodata & \nodata & \nodata & \nodata & \nodata & \nodata & \nodata & \nodata & \nodata &   \nodata  \\
10.0 &    1.52 &    0.08 &    1.19 &    1.25 & \nodata & \nodata & \nodata & \nodata & \nodata & \nodata & \nodata & \nodata & \nodata &   \nodata  \\
10.5 &    1.71 &    0.07 &    1.39 &    1.42 & \nodata & \nodata & \nodata & \nodata & \nodata & \nodata & \nodata & \nodata & \nodata &   \nodata  \\
11.0 &    1.92 &    0.05 &    1.63 &    1.60 & \nodata & \nodata & \nodata & \nodata & \nodata & \nodata & \nodata & \nodata & \nodata &   \nodata  \\
11.5 &    2.13 &    0.04 &    1.76 &    1.90 & \nodata & \nodata & \nodata & \nodata & \nodata & \nodata & \nodata & \nodata & \nodata &   \nodata  \\
12.0 &    2.34 &    0.03 &    1.89 &    2.15 & \nodata & \nodata & \nodata & \nodata & \nodata & \nodata & \nodata & \nodata & \nodata &   \nodata  \\
12.5 &    2.55 &    0.03 &    2.02 &    2.40 & \nodata & \nodata & \nodata & \nodata & \nodata & \nodata & \nodata & \nodata & \nodata &   \nodata  \\
13.0 &    2.77 &    0.02 &    2.17 &    2.64 &    2.56 &    0.11 &    2.25 &    1.00 &    2.28 & \nodata & \nodata & \nodata & \nodata &   \nodata  \\
13.5 &    2.97 &    0.02 &    2.28 &    2.87 &    2.80 &    0.09 &    2.37 &    1.00 &    2.60 &    3.15 &    0.27 &    2.66 &    1.00 &      2.99  \\
14.0 &    3.16 &    0.01 &    2.39 &    3.08 &    2.97 &    0.07 &    2.45 &    1.00 &    2.81 &    3.33 &    0.23 &    2.77 &    1.00 &      3.19  \\
14.5 &    3.34 &    0.01 &    2.47 &    3.28 &    3.20 &    0.06 &    2.53 &    0.96 &    3.11 &    3.41 &    0.21 &    2.87 &    0.95 &      3.28  \\
15.0 &    3.47 &    0.01 &    2.43 &    3.43 &    3.39 &    0.04 &    2.60 &    0.94 &    3.34 &    3.67 &    0.16 &    2.96 &    0.95 &      3.60  \\
15.5 & \nodata & \nodata & \nodata & \nodata &    3.66 &    0.03 &    2.66 &    0.94 &    3.64 &    3.88 &    0.13 &    3.03 &    0.93 &      3.84  \\
16.0 & \nodata & \nodata & \nodata & \nodata &    3.96 &    0.02 &    2.71 &    0.92 &    3.97 &    4.03 &    0.11 &    3.09 &    0.91 &      4.02  \\
16.5 & \nodata & \nodata & \nodata & \nodata &    4.18 &    0.02 &    2.78 &    0.89 &    4.22 &    4.29 &    0.08 &    3.14 &    0.88 &      4.32  \\
17.0 & \nodata & \nodata & \nodata & \nodata &    4.31 &    0.02 &    2.85 &    0.85 &    4.37 &    4.34 &    0.08 &    3.20 &    0.81 &      4.39  \\
17.5 & \nodata & \nodata & \nodata & \nodata &    4.40 &    0.01 &    2.91 &    0.51 &    4.68 &    4.36 &    0.08 &    3.23 &    0.72 &      4.47  \\
18.0 & \nodata & \nodata & \nodata & \nodata & \nodata & \nodata & \nodata & \nodata & \nodata &    4.34 &    0.08 &    3.26 &    0.56 &      4.55  \\
\enddata
\tablenotetext{a}{Star galaxy separation using morphology criterion
(see text).}
\tablenotetext{b}{Star count estimates from the DIRBE Faint Source
Model \citep{dirbe,Wainscoat1992}.}
\tablenotetext{c}{Galaxy counts corrected for incompleteness.}
\tablecomments{Columns labelled ``Total'' tabulate the logarithm of
the observed differential number counts in units of
number~mag$^{-1}$~deg$^{-2}$.  Columns labeled ``$\pm$'' give the
Poisson uncertainty of the total counts in logarithmic
units. ``Star'' counts are in the same units.  For the EGS and QSO1700
fields, the completeness (``Comp.'') estimated from the Monte Carlo
simulations is tabulated and applied to the corresponding ``Galaxy'' counts.
Total source counts are provided in the
Bo\"otes field fainter than stars and galaxies could be reliably
separated using the morphology criterion (see text).  }
\end{deluxetable}

\end{document}